\documentclass[12pt]{iopart}
\usepackage{graphicx}
\usepackage{dcolumn}
\usepackage{bm}
\usepackage[usenames,dvipsnames]{color}
\expandafter\let\csname equation*\endcsname\relax
\expandafter\let\csname endequation*\endcsname\relax
\usepackage{amsmath}
\usepackage{amssymb}

\newcommand{\ket}[1]{|#1\rangle}

\newcommand{\dd}[1]{\frac{\text{d}}{\text{d}t}}

\newcommand{\roughly}{$\sim$} % μm

\newcommand\nbar{\bar{n}}

\newcommand{\cSig}{c^3\Sigma_1}

\newcommand{\XSig}{X^1\Sigma}

\newcommand{\um}{\text{\textmu}{\rm m}}
\newcommand{\us}{\text{{\textmu}s}}

\begin{document}

\title[An optical tweezer array of ground-state polar molecules]{An optical tweezer array of ground-state polar molecules}

\author{Jessie~T.~Zhang$^{1,2,3}$, Lewis~R.~B.~Picard$^{1,2,3}$, William~B.~Cairncross$^{2,1,3}$\footnote{Present address: Atom Computing Inc, Boulder, CO 80301, USA}, Kenneth~Wang$^{1,2,3}$,  Yichao~Yu$^{1,2,3}$\footnote{Present address: Duke Quantum Center, Durham, NC 27701, USA.}, Fang~Fang$^{2,1,3}$, Kang-Kuen~Ni$^{2,1,3}$}

\address{$^1$ Department of Physics, Harvard University, Cambridge, Massachusetts 02138, USA}
\address{$^2$ Department of Chemistry and Chemical Biology, Harvard University, Cambridge, Massachusetts 02138, USA}
\address{$^3$ Harvard-MIT Center for Ultracold Atoms, Cambridge, Massachusetts 02138, USA}
\ead{\mailto{jessiezhang@g.harvard.edu}, \mailto{ni@chemistry.harvard.edu}}
\vspace{10pt}

\begin{abstract}
Fully internal and motional state controlled and individually manipulable polar molecules are desirable for many quantum science applications leveraging the rich state space and intrinsic interactions of molecules. While prior efforts at assembling molecules from their constituent atoms individually trapped in optical tweezers achieved such a goal for exactly one molecule \cite{Zhang2020, Cairncross2021, He2020}, here we extend the technique to an array of five molecules, unlocking the ability to study molecular interactions.  We detail the technical challenges and
solutions inherent in scaling this system up. With parallel preparation and control of multiple molecules in hand, this platform now serves as a starting point to harness the vast resources and long-range dipolar interactions of molecules. 
\end{abstract}

\section{Introduction}

Optical tweezer arrays of neutral atoms has emerged as a powerful platform for quantum simulation, information, and metrology applications with their single particle control, configurability and scalability \cite{Browaeys2020, Kaufman2021}. With the addition of Rydberg excitations, atoms can interact via strong dipolar interactions over tens of micrometer separations, allowing interaction ranges to be tuned from pairs of atoms to the complete system size. These systems have recently been used to study phase transitions and new phases of matter \cite{Bernien2017,Scholl2021,Semeghini2021,Choi2021}, achieve half-a-minute coherence time of neutral atom optical clocks \cite{Young2020}, and demonstrate great promise in  quantum computing applications \cite{  Graham2019,Omran2019, Madjarov2020}.  

Compared to their neutral atom counterparts, ultracold polar molecules possess a richer set of internal structures and can exhibit intermolecular dipole-dipole interactions~\cite{Carr2009}.
These intrinsic coherent features offer even more opportunities to tailor molecular quantum systems to specific applications, including studying quantum many-body physics~\cite{Micheli2006, Gorshkov2011a, Yao2018} and engineering robust storage and transmission of quantum information~\cite{DeMille2002,  Ni2018, Park2015,gregory_robust_2021, Burchesky2021}. However, reaching a high-level of control similar to what has been achieved with atoms is a desirable starting point for many applications. There have thus far been two approaches to preparing single molecules in optical tweezers. In one approach, molecules can be associated from single atoms trapped in optical tweezers~\cite{Liu2018}. Full internal and motional quantum state control are inherent in the molecule creation process and have been achieved for a rovibrational ground state NaCs molecule, but the demonstration has so far been limited to creating a single molecule at a time \cite{Cairncross2021}. In another approach, molecules can be directly laser cooled and trapped in optical tweezers \cite{Anderegg2019}. The demonstration has shown to be parallelizable for a tweezer array of CaF molecules, but these molecules are not yet cooled to the motional ground state \cite{Caldwell2020_cooling}. 

\begin{figure}
\centering
\includegraphics[width=0.8\textwidth]{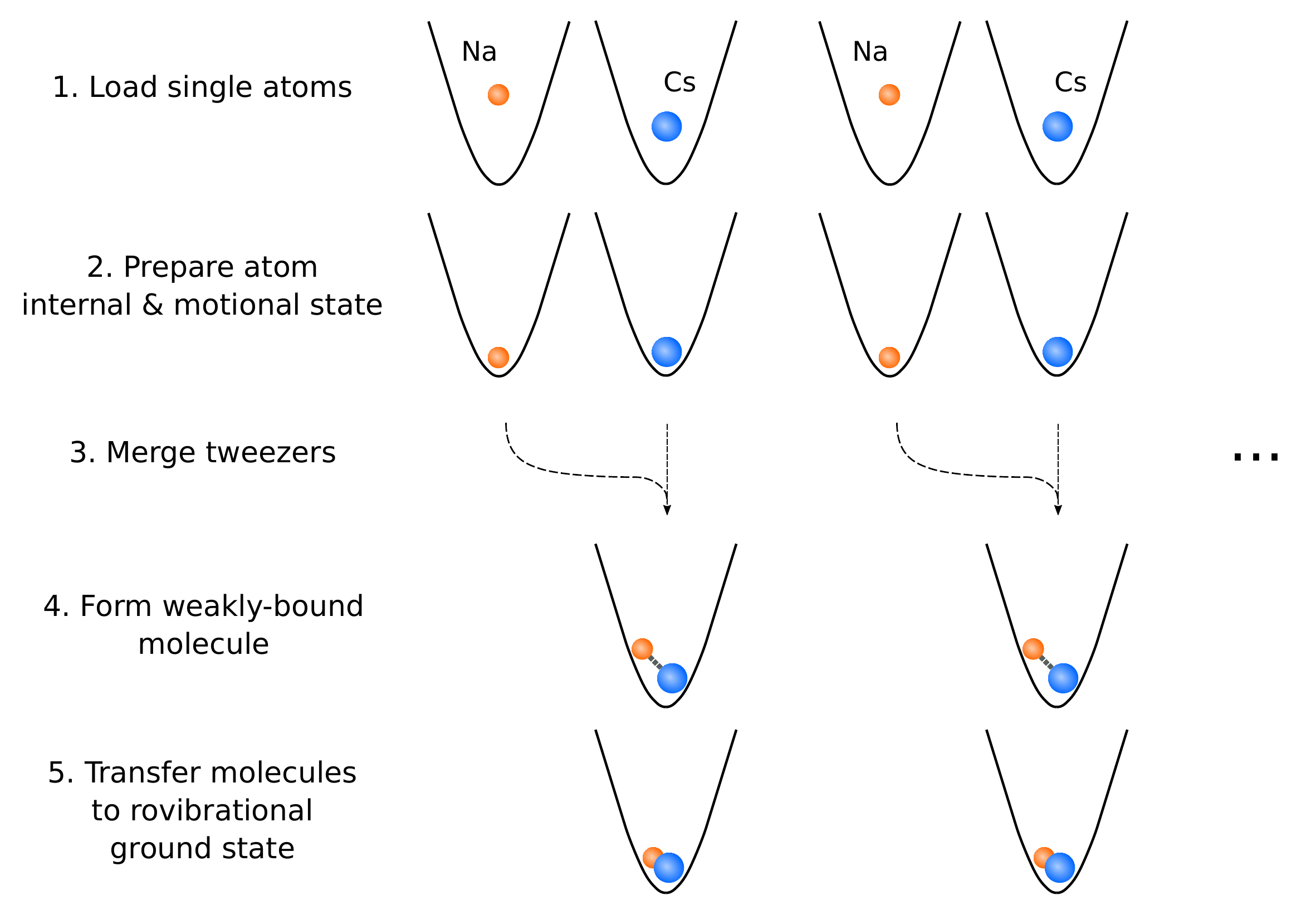}
\caption[Molecule array formation scheme]{\textbf{Molecule array formation scheme}. Arrays of atom pairs are associated coherently into rovibrational ground state molecules in parallel to form optical tweezer arrays of ground-state molecules.
\label{fig:array-formation-schematic}}
\end{figure}

In this work, we use the former approach to demonstrate the creation of an optical tweezer array of rovibrational ground state molecules with full internal and motional quantum-state control. We choose NaCs as our molecular species for its large molecular frame dipole moment (4.6 Debye), which is advantageous in applications leveraging their strong dipolar interactions.
We assemble molecules from single atoms following the steps shown schematically in Fig.~\ref{fig:array-formation-schematic}. Different from our prior work \cite{Zhang2020,Cairncross2021}, we perform these steps in parallel starting with multiple tweezers of each species. Specifically, we trapped 1D arrays of 5-10 Na and Cs atoms side-by-side and cooled the individual atoms to their motional ground state simultaneously. We then merge pairs of traps adiabatically to prepare a single Na and a single Cs atom in the same trap. The atom pairs are then converted adiabatically to a single molecule by a two step process: first magnetoassociating to a weakly-bound Feshbach molecule, then transferring to the rovibrational ground state by a coherent two-photon process. 

In this paper, we detail the technical challenges to scaling the system up and the solutions we implemented to overcome them. Our findings are applicable to many quantum science platforms seeking to scale up in sizes, and provides a starting point for quantum science studies using ultracold molecules in optical tweezers.

\section{Dual species array}

\subsection{Atom loading}
The experiment starts with a dual species magneto-optical trap consisting of approximately $10^5$ $^{23}$Na and $^{133}$Cs atoms each. Single Na and Cs atoms are then loaded stochastically by means of a parity-selection process from light-assisted collisions \cite{Schlosser2001} into separate 1D arrays of optical tweezers at 623~nm and 1064~nm, respectively. This wavelength choice allows for independent control over the two species, which is crucial for selectively loading the two species into their respective traps and adiabatically merging the two traps prior to molecule formation \cite{Liu2019}. This necessitates individual control over the two optical tweezer beampaths, which we detail below.

\subsubsection{Optical setup}

\begin{figure}
\centering
\includegraphics[width=0.8\textwidth]{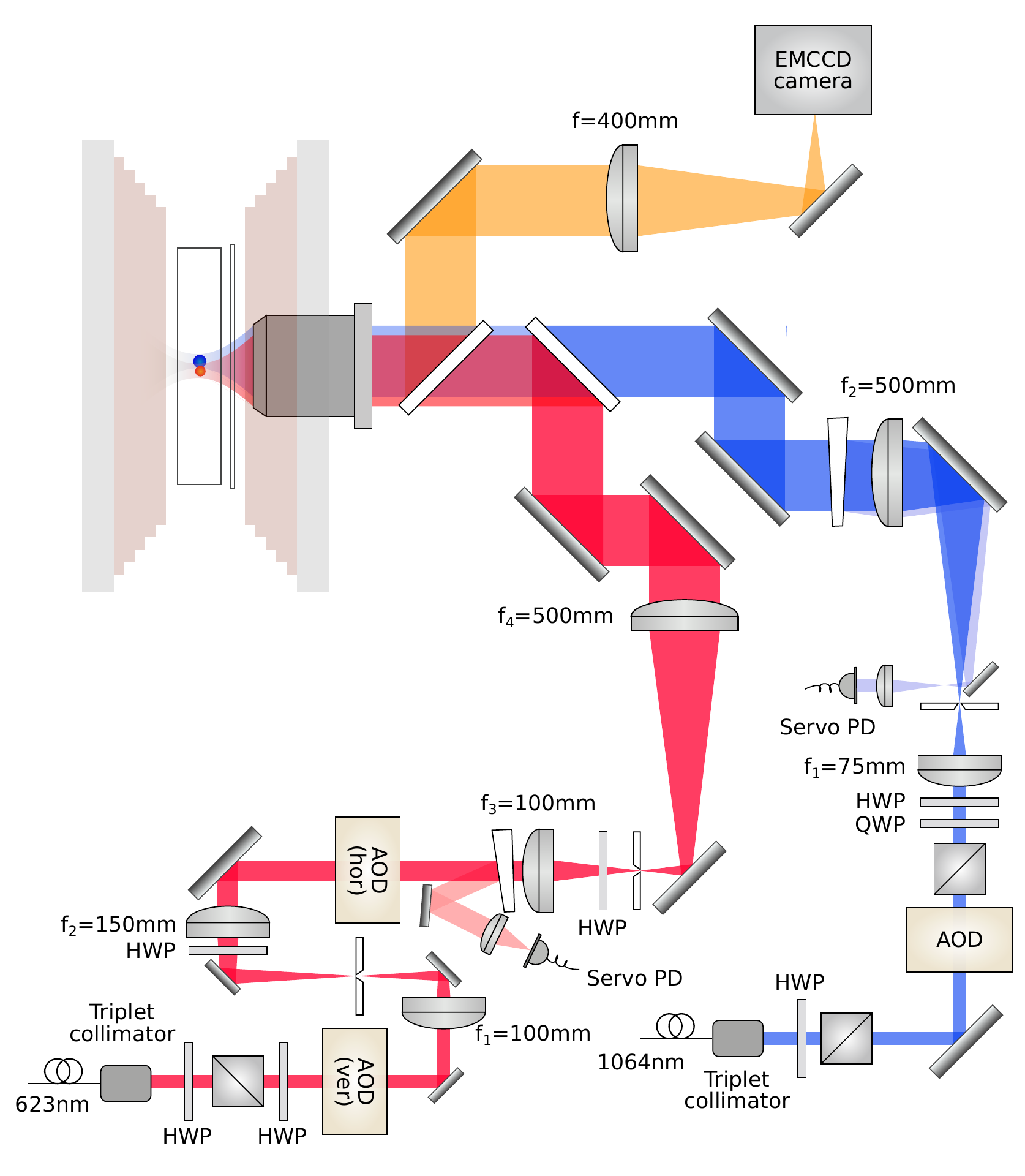}
\caption[Schematic of tweezer beampath]{\textbf{Schematic of tweezer beampath}. Tweezer beams for Na (623~nm) and Cs (1064~nm) are launched individually and sent through AOD's to generate multiple beam deflections that correspond to multiple optical tweezer sites at the location of the atoms. The tweezer intensities are actively stabilized after the AOD's picked off from a wedge in the beampaths. (Not drawn to scale).
\label{fig:beampath-tweezers-arrays}}
\end{figure}

A schematic of the optical tweezer beampaths is shown in Fig.~\ref{fig:beampath-tweezers-arrays}. To generate an array of optical tweezer beams, we use longitudinal-mode acoustic-optical deflectors (AOD) which provide multiple deflections when driven with multiple radio frequency (RF) tones simultaneously. A single AOD is used in the 1064~nm beampath to generate a 1D array of beams, while two orthogonal ones are placed in the 623~nm beampath to also allow merging the two parallel 1D arrays. Notably, this configuration allows for \emph{in situ} rearrangement~\cite{Endres2016, Barredo2016} to create defect-free arrays in future work. The 623~nm and 1064~nm beams are combined before entering the objective (NA=0.55) which produce the optical tweezer arrays at the center of the focus. Fluorescence from the atoms is then collected through the same objective and imaged on an EMCCD camera. 

To generate the high laser power necessary for trapping and cooling Na atoms, we use a MgO:PPLN sum-frequency generation crystal to produce $623$~nm from $1038$~nm and $1550$~nm sources. We optimize the input beam sizes according to the Boyd-Kleimann focusing condition \cite{Boyd1968}, and obtain \roughly$5$~W of 623~nm light after the crystal with 9.1~W and 8.6~W input of 1038~nm and 1550~nm respectively. 

\subsubsection{AOD RF control}

We use an arbitrary waveform generator (AWG) to simultaneously generate multiple RF tones for the AOD's in the horizontal direction for the 623~nm and 1064~nm beampaths, where the amplitude and phase for each tone can be individually modified. We initially found tweezer-induced scrambling of $m_F$ states of the atoms when operating at low amplitudes, likely due to noise from the AWG which gets amplified in the RF circuit downstream; we therefore opt to operate at maximum amplitudes of the AWG to eliminate such $m_F$ scrambling. 

The RF signal output from the AWG is nominally of the form 
\begin{equation}
    \sum_i\cos\left(2\pi f_i t + \phi_i\right)
\end{equation}
However, due to non-linearities in the system, whether from the AWG, RF amplifiers, or the AOD, terms in this sum can mix and result in terms
\begin{equation}
    \cos(2\pi f_i t + \phi_i)\cos(2\pi f_j t + \phi_j)\propto\cos\left(2\pi(f_i-f_j)+(\phi_i-\phi_j)\right)+\cos\left(2\pi(f_i+f_j)+(\phi_i+\phi_j)\right)
\end{equation}
The terms involving $f_i-f_j$ can then mix again with the original signal $f_k$ and result in terms near the original desired tones $f_k + (f_i-f_j)$. For a uniformly spaced array, where the neighboring $f_i$ are equally spaced, these terms can collude with the original signal and  interfere leading to unequal amplitudes in the desired tones. There are methods to optimize this to avoid interference effects based on a global search of optimized phases and amplitudes of the individual tones \cite{Endres2016}. In practice, we found that by starting from a set of phases randomly distributed between 0 and $2\pi$ for the different tones, destructive interference was not significant, and we could optimize homogeneity of the traps by adjusting amplitudes only in the experiment. We use the trapping frequencies of the traps on the atoms as the metric to fine-tune the homogeneity of each array.

\subsubsection{Heating problems}

\begin{figure}
\centering
\includegraphics[width=0.6\textwidth]{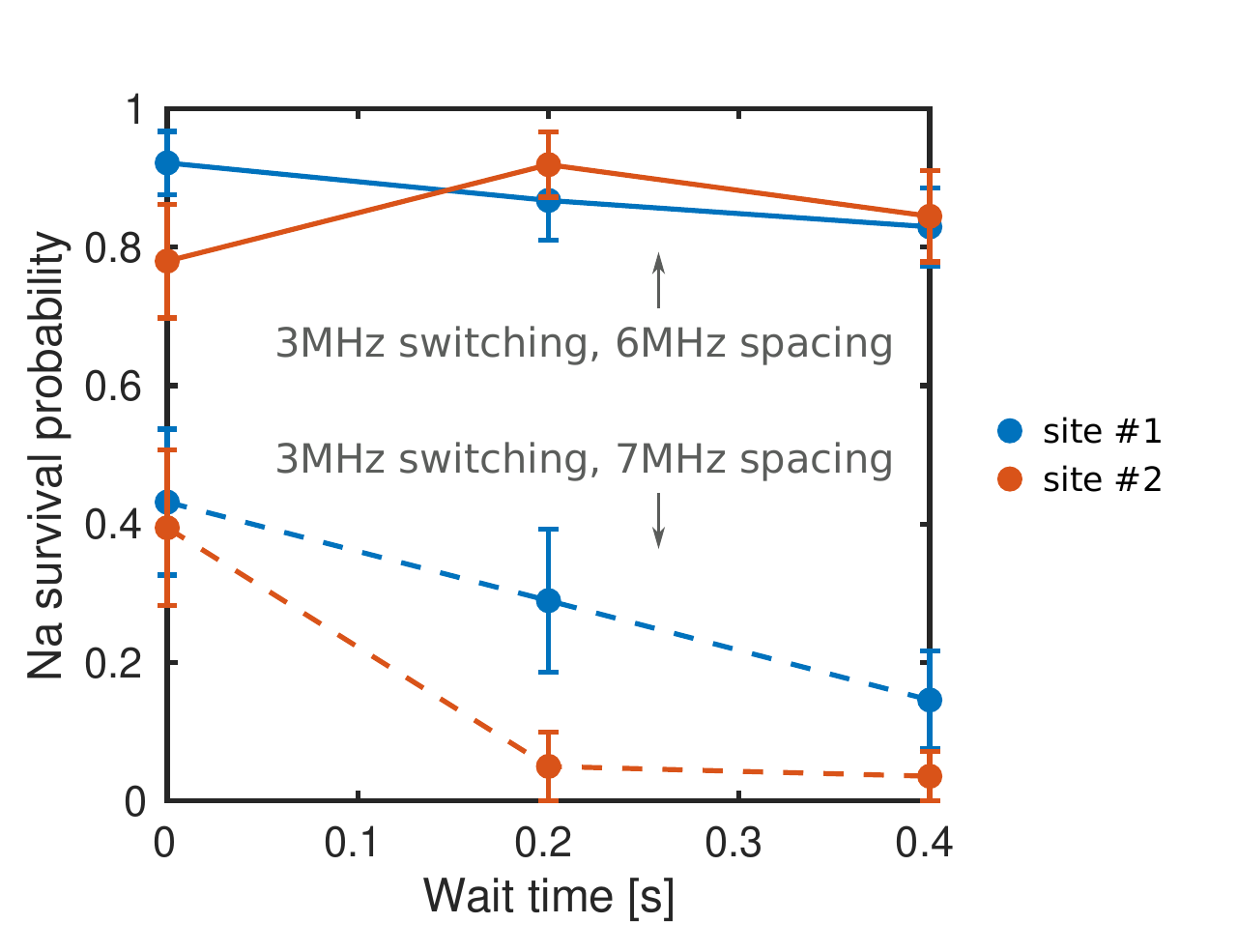}
\caption[Heating of Na atoms in array]{\textbf{Heating of Na atoms in array.} The tweezer and imaging light is pulsed out of phase at 3~MHz. Solid (dashed) line shows survival of two Na atoms spaced 6~MHz (7~MHz) apart and held at the same trap depth (\roughly 3mK). In particular, we see significant heating and loss when the AOD frequency spacing is not a multiple of the switching frequency.
\label{fig:na-heating}}
\end{figure}

In the experiment, we modulate the Na trap light and the laser cooling/imaging light in square pulses out of phase with each other to eliminate light shifts from the tweezer light that inhibits loading and imaging \cite{Hutzler2017}. In the array, we found empirically that when the array spacing frequency difference is a non-integer multiple of the modulation frequency, excessive heating would occur. An example of this is shown in Fig. \ref{fig:na-heating} where we load and measure the lifetime of the Na atoms in two neighboring traps. In particular, when the switching frequency is at 3~MHz, and two traps are spaced 7~MHz away, rapid loss was seen from the trap. However, with a spacing of 6~MHz (an integer multiple of 3~MHz), no significant heating was observed.

The switching of the laser frequency $f_L$ at $f_S$ with a 30\% duty cycle square pulse leads to frequency sidebands on the laser at frequencies $f_L+n\cdot f_S$, where $n$ is any integer. In the simple case of two tones $f_1$ and $f_2$ driving the AOD, the laser beam is deflected and generates two beams containing frequency components
\begin{equation}
    f_L+n\cdot f_S+f_1 \text{  and  } f_L+m\cdot f_S+f_2
\end{equation}
If the two beams overlap in space at the location of the atoms, this can lead to beating at frequencies
\begin{equation}
    (f_1-f_2) - n\cdot f_S
\end{equation}
In particular, when $\Delta f = f_1-f_2$ is not an integer multiple of $f_S$, this difference can lead to frequencies near two times the trapping frequency, causing parametric heating on the atoms. Therefore, we found it necessary to find frequencies that avoid this situation.

Our choice of switching frequency $f_S$ is limited by various considerations \cite{Hutzler2017}. On the one hand, the switching frequency needs to be less than the excited state linewidth (\roughly$2\pi\times10$~MHz) so that the atom has sufficient time to scatter light to enable laser cooling. At the same time, the switching frequency needs to be larger than the trapping frequency (\roughly500~kHz) so that the atom motion in the trap is unaffected by the switching. This constrains our choice of frequency spacing for the AOD array. We choose a switching frequency $f_S$ of 3.5~MHz and AOD frequency spacing $\Delta f$ of 7~MHz or 14~MHz for the Na array. To align the Na and Cs traps, this corresponds to an AOD frequency spacing of 5.28~MHz or 10.57~MHz for Cs.

\subsubsection{Array loading}

\begin{figure}
\centering
\includegraphics[width=0.95\textwidth]{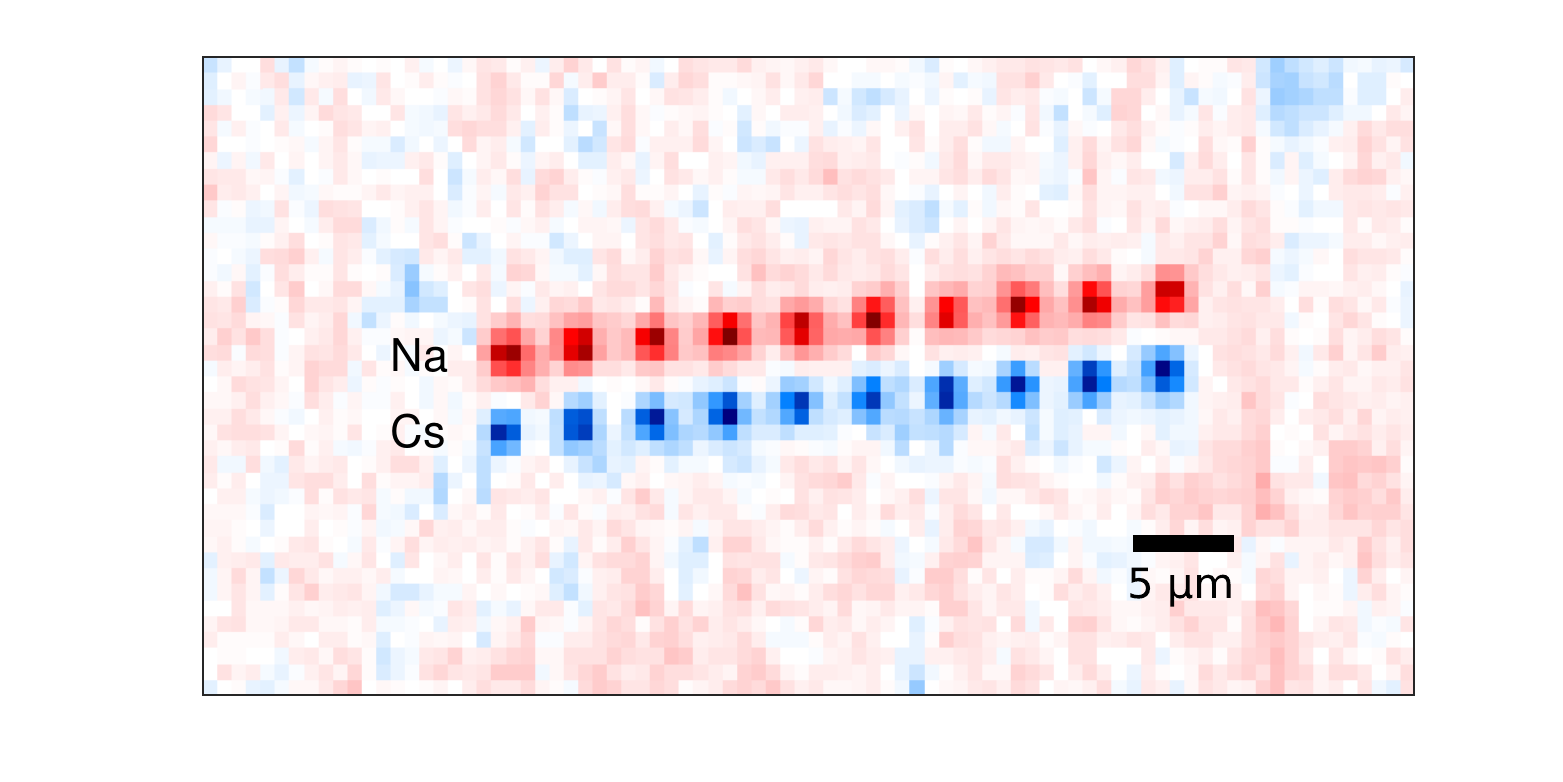}
\caption[Dual species array of single atoms.]{\textbf{Dual species array of single atoms.} Na and Cs arrays are generated individually by AOD's in the horizontal direction. Na and Cs atoms are imaged separately and overlaid to produce the image, which is averaged over 132 experimental runs. Red (blue) corresponds to trapped single Na (Cs) atoms. The atoms are spaced \roughly3.7~\um~in their individual arrays, while the spacing between the Na and Cs arrays is \roughly3~\um. A separate AOD in the vertical direction for the 623~nm tweezer steers the Na array to merge with 1064~nm Cs tweezer beams.
\label{fig:nacs-array}}
\end{figure}

In Fig. \ref{fig:nacs-array} we show a dual species array consisting of 10 Na and Cs atoms each. The image is averaged over 132 experimental cycles. In the experiment, the Na and Cs atoms are imaged sequentially to reduce background, and the images are overlaid to show two species simultaneously. The atoms within each row are spaced by 3.6~\um, and the two rows are spaced by 3.3~\um, corresponding to 7~MHz and 5.28~MHz spacing for the 623~nm and 1064~nm horizontal AOD's respectively. We achieve 50.3(4)~\% and 62.6(4)~\% loading rates for Na and Cs respectively averaged over the individual sites, comparable to the loading rate achieved in single tweezers.

\subsection{Raman sideband cooling}

\begin{figure}
\centering
\includegraphics[width=0.8\textwidth]{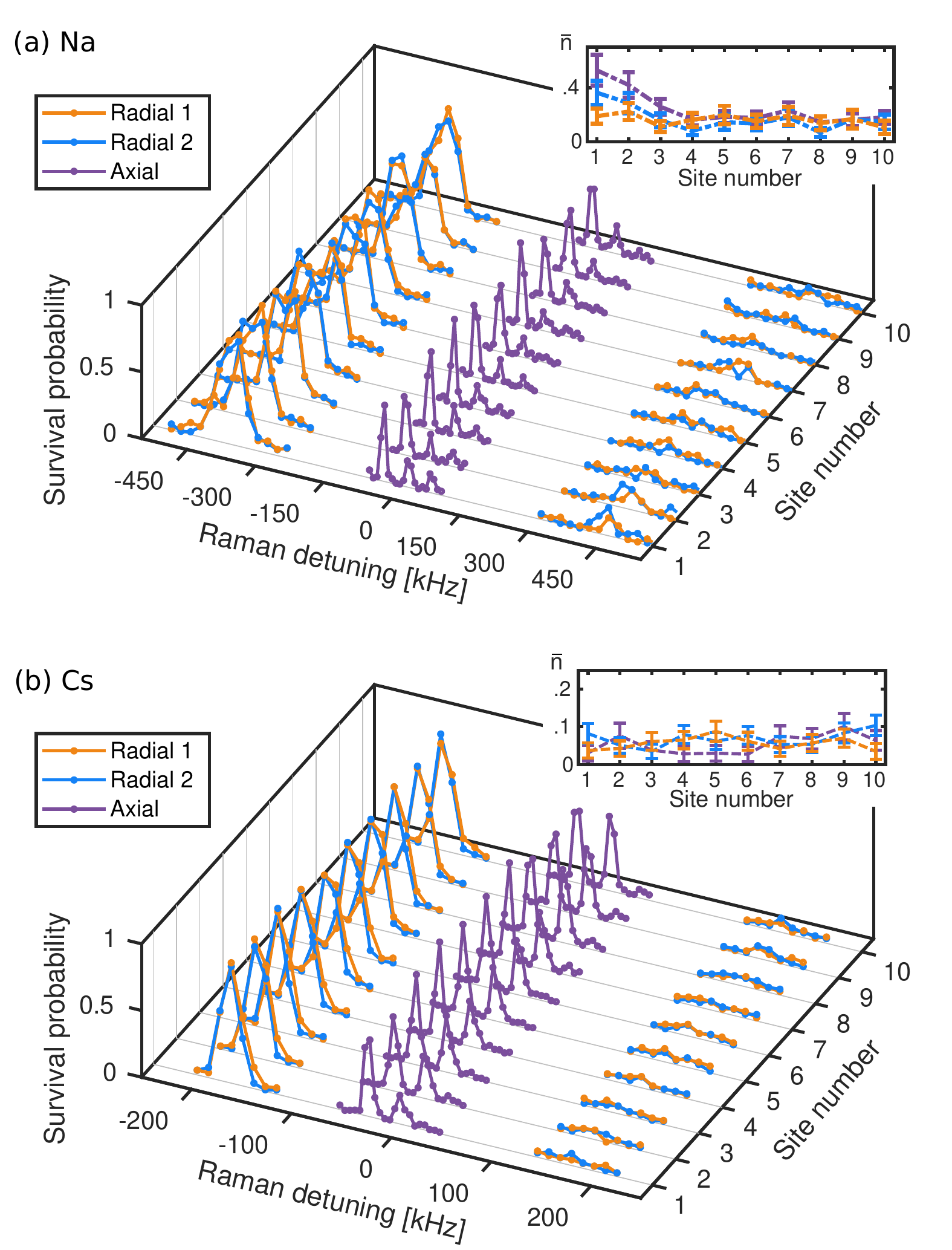}
\caption[Raman sideband thermometry in an array after cooling]{\textbf{Raman sideband thermometry in an array after cooling.} 10 atoms of each species are loaded simultaneously and Raman sideband cooled. Raman sideband thermometry is performed after the cooling sequence, and is shown for (a) Na and (b) Cs individually. Insets: Site-by-site average motional quanta $\nbar$ values after cooling for each of the 3 axes.
\label{fig:array-rsc}}
\end{figure}

After we have trapped a dual species array, we Raman sideband cool the array of atoms to their 3D motional ground states simultaneously. This is necessary for coherently forming molecules by magnetoassociation in the following step. The cooling sequence we use is similar to that discussed in our previous work \cite{Yu2018,Liu2019}, which alternates cooling pulses between the two radial and axial axes. The cooling process takes 60~ms in total. The Raman beams are \roughly200-300~\um~in diameter, which is sufficient to address the entire array homogeneously with proper centering of the beams on the array.

In Fig. \ref{fig:array-rsc} we show Raman sideband thermometry spectra obtained after Raman sideband cooling of an array of 10 Na (a) and Cs (b) atoms simultaneously. The mean trapping frequencies are $\omega_\text{Na} = 2\pi \times (369,371,54)$~kHz and $\omega_\text{Cs} = 2\pi \times (168,170,34)$~kHz respectively, with $<$3\% variation across the array. In the insets we show the average motional quanta $\nbar$ for each of the 3 axes for each species at each individual site after cooling. On average, we achieve $\nbar_\text{Na} = (0.15(5),0.13(5),0.18(4))$ and $\nbar_\text{Cs}=(0.056(22),0.068(23),0.054(25))$ over all sites. This presents a starting point for many dual species applications where motional control of the atoms is crucial, including molecule formation which we detail below.

\section{Molecule formation}

\subsection{Forming Feshbach molecules}

With the atoms Raman sideband cooled, we proceed to form Feshbach molecules using the procedure described in Ref. \cite{Zhang2020}. We first ramp the magnetic field to 867~G using a pair of large coils driven in Helmholtz configuration before merging the traps so that a single pair of Na and Cs atoms occupies each site of the 1064~nm optical tweezer array. The motion is controlled by ramping the frequency of the vertical AOD in the 623~nm tweezer beampath following the sequence described in Ref. \cite{Liu2019}.  The magnetic field is then ramped adiabatically across an s-wave Feshbach resonance located near 864~G which magnetoassociates the atom pair. 

We initially however, found no Feshbach molecule creation, and traced this to loss of $m_F$ population during the magnetic field ramp up to high fields. Specifically, we found, for example, that the population for Cs would be lost in an array spaced by $\Delta f=$5.28~MHz for the 1064~nm array upon going above \roughly15~G. At 15.08G, the neighboring $m_F$ level energy spacing coincides with the neighboring trap frequency difference $\Delta f$. This $m_F$ depumping arises from overlap of neighboring traps generated by the AOD's at the frequency difference $\Delta f$. In the presence of circular polarization of the tweezer beam at location of the atoms, overlapping beams whose frequency difference is on resonance with the $m_F$ transition causes optical Raman transitions between the $m_F$ levels.
An alternative way of understanding this is that the atoms see an effective magnetic field at the center of the tight optical tweezer due to the breakdown of the paraxial approximation \cite{Kaufman2012, Thompson2012}. Optical beating causes this effective magnetic field to oscillate, which in turn causes $m_F$ transitions that scrambles the internal state when on resonance. The overlap of neighboring beams is exacerbated by any clipping in the beampath, which causes Airy disk patterns in the images of the individual trap beams. 

\begin{figure}
\centering
\includegraphics[width=\textwidth]{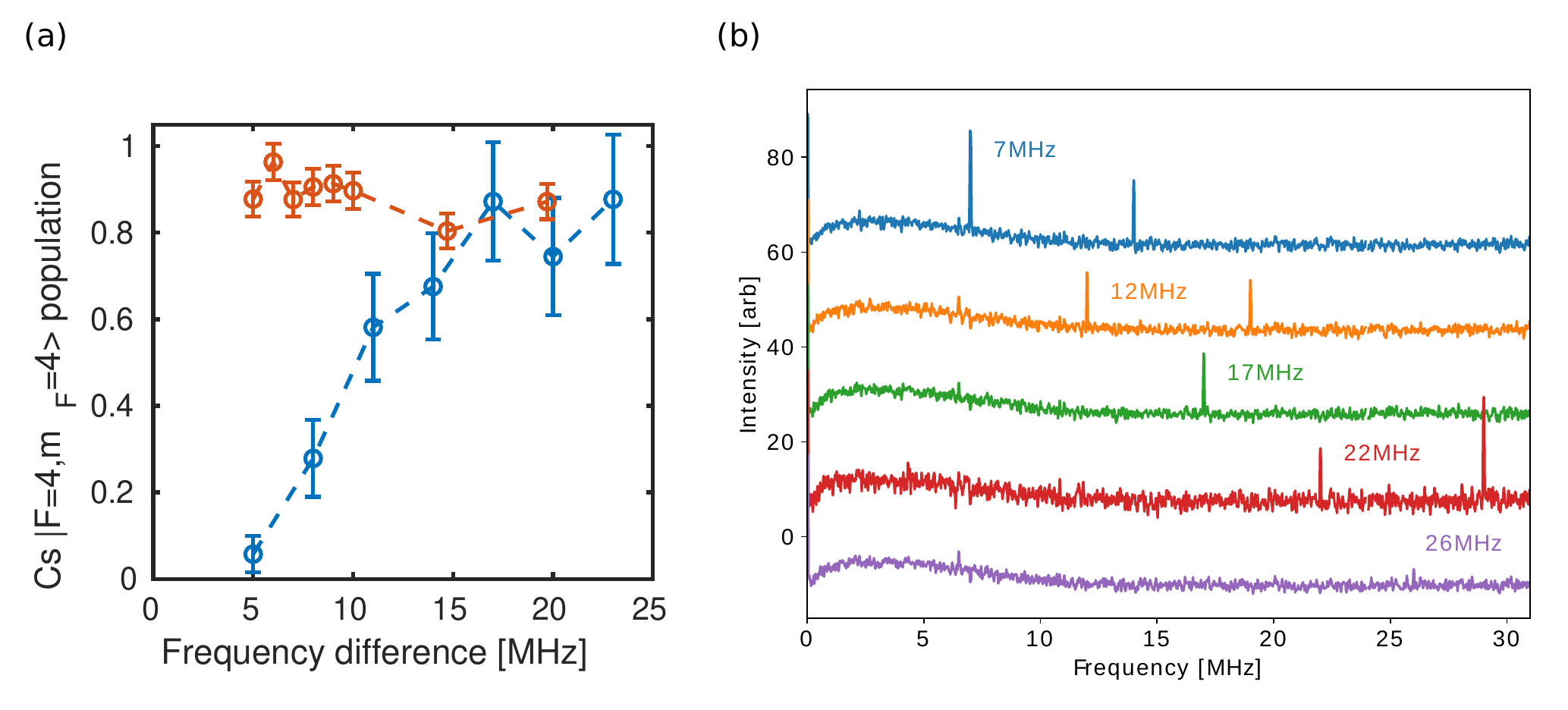}
\caption[$m_F$ depumping of Cs atom in an array]{\textbf{$m_F$ depumping of Cs atom in an array.} (a) Depumping of a single Cs atom out of the stretched state $\ket{F=4,m_F=4}_\text{Cs}$ with varying neighboring AOD trap frequency difference. Blue (red) points indicate the depumping before (after) implementing the changes discussed in the text to mitigate depumping. (b) Optical intensity noise spectrum of a single trap measured on a photodiode with varying neighboring trap frequency spacing. The vertical axis is offset for clarity. The tweezer beam is clipped at the focus to pick out a single trap.
\label{fig:fb-mol-beating}}
\end{figure}

In Fig. \ref{fig:fb-mol-beating}(a) we show the depumping of a single Cs atom from the stretched state $\ket{F=4,m_F=4}_\text{Cs}$ as a function of neighboring trap frequency. The blue data points show the case of bad depumping, whereby traps spaced closer than 15~MHz in driving frequency corresponding to 9.3~\um~causes depumping out of the stretched state. In Fig. \ref{fig:fb-mol-beating}(b) we measure the intensity spectrum of a single trap clipped off at the focus. We find a decreasing beatnote as the neighboring trap frequency difference is increased, consistent with the falloff in depumping effect. To mitigate these effects, we jump the magnetic field from 0 to $80$~G which settles in 200\us~before ramping the coils up to the desired field. We also ramp the trap intensities down to 20 kW/cm$^2$, which corresponds to \roughly 5\% the loading trap depths, during the magnetic field ramp, which also lowers the depumping rate. The depumping after these implementations are shown in red in Fig. \ref{fig:fb-mol-beating}(a). While this effectively eliminates the depumping effects, we opt to separate the traps to double the spacing, so that the frequency is 14~MHz and 10.57~MHz for Na and Cs respectively, corresponding to a distance of 7.4~\um, to preempt further problems during molecule formation. Thus, while the optical system is capable of resolving 10 or more traps in one dimension, in the present work we demonstrate 5 traps for rovibrational ground state formation in the next section.

\subsection{Array of rovibrational ground state molecules}

In Ref. \cite{Cairncross2021} we identified a two-photon pathway to transfer NaCs Feshbach molecules to the rovibrational ground state $\ket{\XSig, v=0, N=0}$ using the intermediate state $\ket{\cSig, v'=26, J'=1, m_J'=1}$. Notably, we found that it was more straightforward, compared to the established method of STIRAP in the literature for associating bi-alkali species \cite{STIRAP}, to apply a detuned Raman pulse to transfer the population due to the large excited state scattering rate. While this method mitigates complications arising from the large scattering rate, it also places more stringent requirements on the intensity stabilization and uniformity of the addressing beams and thus presents challenges when scaling up to an array of molecules. In the present work, we optimized the Raman transfer parameters to minimize the sensitivity of AC Stark shift on beam intensities, and uniformly address the atoms in the array by creating a top-hat beam shape.

\subsubsection{Uniformly addressing arrays}

\begin{figure}
\centering
\includegraphics[width=0.95\textwidth]{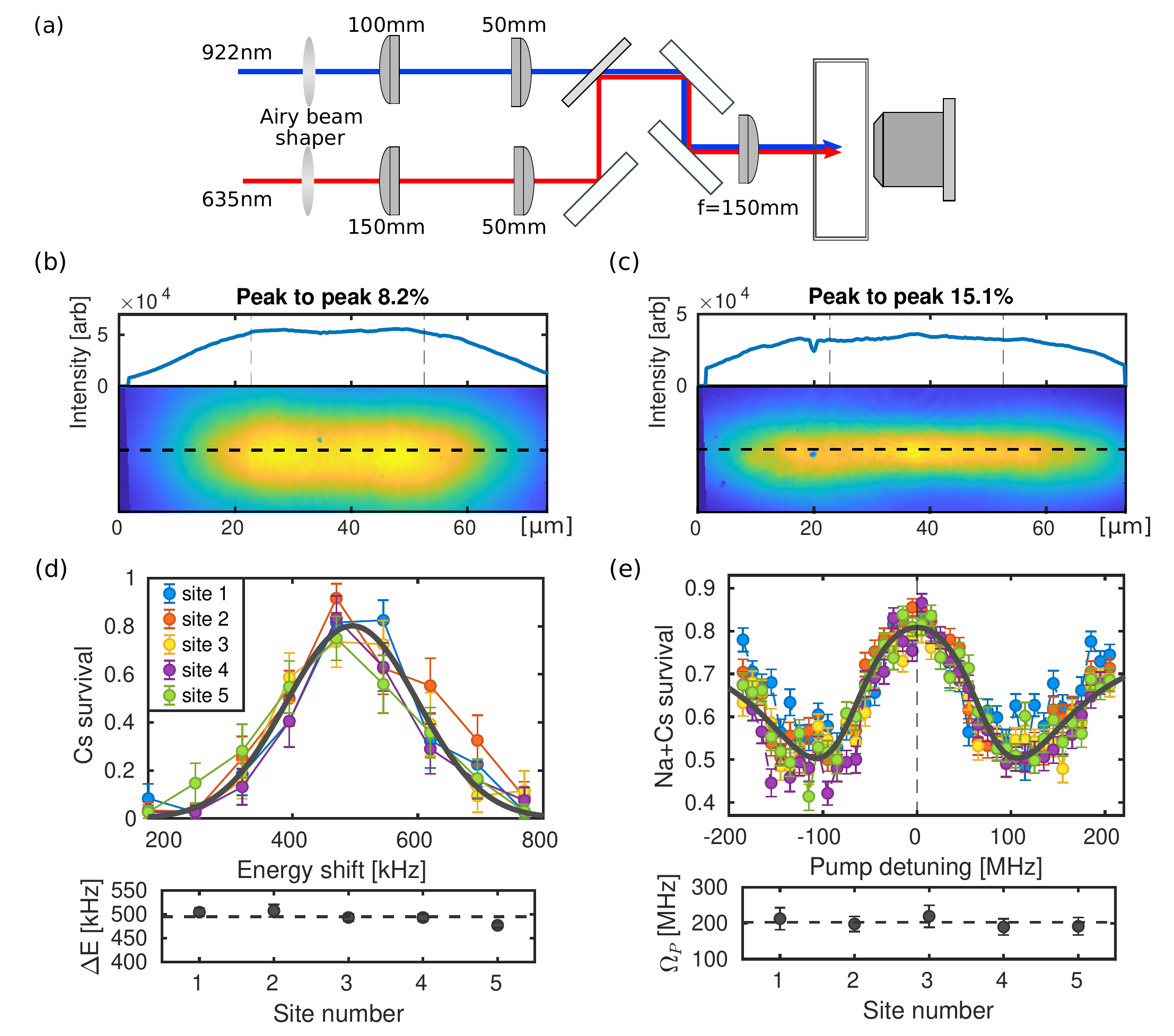}
\caption[Tophat beam shaping]{\textbf{Tophat beam shaping.} (a) Schematic of tophat beampath. (b-c) The intensity profiles of the (b) 922~nm and (c) 635~nm beams are imaged on a beam profiler at the focus of the beam. The vertical direction is a Gaussian profile. The tophat profile is in the horizontal direction to address the array. The peak-to-peak variation is taken over 30~\um, which is the array size to address. (d) AC Stark shift measurement to fine-align the pump beam. (e) Autler-Townes spectroscopy measurement to fine-align the Stokes beam.
\label{fig:tophat-beam}}
\end{figure}

To uniformly address the array with the Raman transfer beams, we create a tophat beam shape that spans the 1D array. We use an Airy beam shaper (Asphericon ASM25-10-D-B-780 and ASM25-10-D-B-632 for 922~nm and 635~nm respectively) that forms an Airy disk phase pattern in the Fourier plane, which then gets imaged by a lens to form a tophat beam shape at the location of the atoms. We use an elliptically shaped beam to address the 1D array for higher intensities. A schematic of the beam path is shown in Fig. \ref{fig:tophat-beam}(a). The beam shape is sensitive to the alignment along the z-direction, as the beam profile varies drastically around the focal plane. Since the beam is counter-propagating to the optical tweezer beam path, we image the beam on the EMCCD camera we normally use to image the single atoms for alignment in z, and alignment in x-y to the atoms/molecules. We are able to achieve an intensity uniformity of 8\% and 15\% over the peak of the 922~nm (pump) and 635~nm (Stokes) beam respectively on a beam profiler as shown in Fig. \ref{fig:tophat-beam}(b-c).

We use a vector light shift measurement to fine-align the 922~nm beam to the atoms as shown in Fig. \ref{fig:tophat-beam}(d). In particular, the vector light shift between the Cs states $\ket{F=4, m_F=4}$ and $\ket{F=3, m_F=3}$ from a circular polarized beam at 922~nm is proportional to beam intensity \cite{le2013dynamical}. Using this, we find a uniformity of the pump beam of 6(2)\%. To check the uniformity of the 635~nm beam, we use an Autler-Townes spectroscopy measurement \cite{Cairncross2021}. In this measurement, the Stokes beam is tuned on resonance, while the frequency of the pump beam is scanned. We observe an Autler-Townes splitting, which is proportional to the Stokes beam Rabi frequency. This is shown in Fig. \ref{fig:tophat-beam}(e). We find that the uniformity of the 635~nm Rabi frequency is 14(10)\%.

\subsubsection{Parameter selection}\label{sec:raman_params}
~\\
Along with the local intensity of the Raman transfer beams, the overlap of the two-photon Raman resonance location for the different array sites is affected by the single-photon detuning from the intermediate state and the power balance of the two beams. In our previous work we applied the pump and Stokes beams on two-photon resonance and with one-photon detuning 21~GHz red of the intermediate state to coherently transfer the Feshbach molecule population to the rovibrational ground state. In the present work, we reduce the single-photon detuning to 7 GHz. This choice of detuning increases the ratio of the two-photon resonance width to the light shift of the resonance position induced by the pump beam, which lessens the sensitivity of the Raman transfer in the array to pump beam uniformity at the expense of a larger scattering rate. This light shift is dominated by the interaction of the pump beam with the $J' = 2$ manifold of the intermediate vibrational state, which begins 6~GHz above the $\ket{\cSig, v'=26, J' = 1, m_J' = 1}$ intermediate state \cite{Cairncross2021}. While the closer detuning increases the sensitivity to the Stokes beam light shift, we use this to our advantage as it allows us to tune the overlap of the frequency overlap of the two-photon resonance between all array sites using the relative power balance of the pump and Stokes beam. By empirical tuning this power balance, as well as the fine alignment of the beams using a picomotor controlled mirror mount, we are able to overcome the remaining non-uniformity of the top-hat beams to drive a simultaneous Raman $\pi$-pulse across the entire array.

\subsection{Coherent rovibrational ground state molecule formation}

\begin{figure}
\centering
\includegraphics[width=0.7\textwidth]{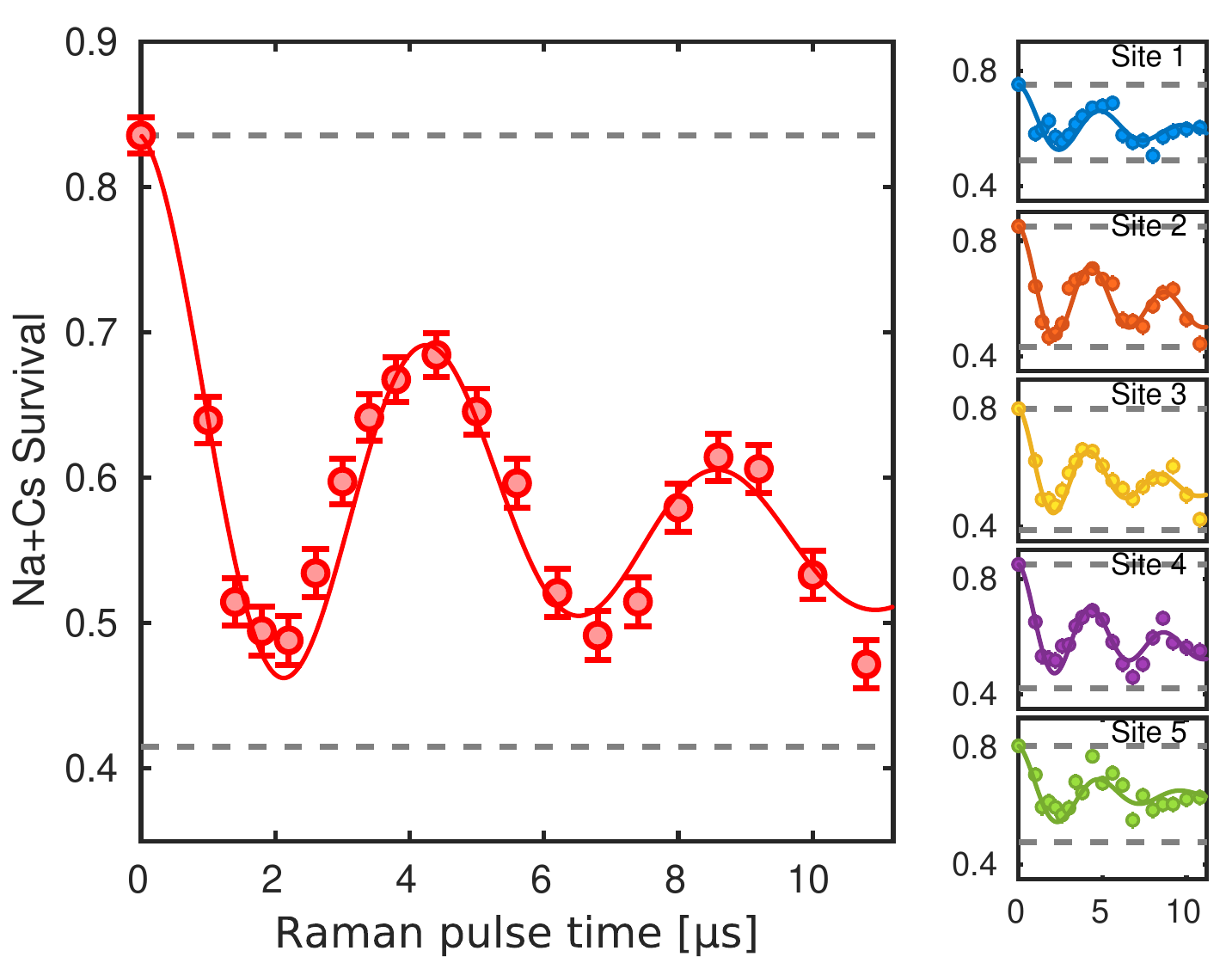}
\caption[Coherent transfer of an array of Feshbach molecules to the rovibrational ground state]{\textbf{Coherent two-photon detuned Raman transfer of an array of Feshbach molecules to the rovibrational ground state.} Left: Average data over 5 sites. Solid line is fit to an exponentially decaying sinusoidal function. Upper (lower) dashed gray line indicates the baseline two-body survival (baseline Feshbach molecule population). Right: Site-by-site data to show the uniformity across the array.
\label{fig:array-rabi-flopping}}
\end{figure}

Using the parameters above, we start from individual arrays of 5 single Na and Cs atoms to create an array of 5 rovibrational ground state molecules. The transfer from Feshbach molecules to the rovibrational ground state is shown in Fig. \ref{fig:array-rabi-flopping}. For detection, we reverse the magnetoassociation process and separate the traps adiabatically to image the atoms. Rovibrational ground state molecules are not transferred back to atoms through the reverse magnetic field ramp, and therefore ground state molecule production can be indicated by correlated atom pair loss. We use 50~mW for the pump beam 922~nm, and 6~mW power for the Stokes beam 635~nm out of the fiber and achieve a $\pi$-pulse time of 2.16(3)~\us. In this work, the Raman transfer beams are individually intensity stabilized to better than 2\%. The Rabi flopping from Feshbach molecules to the rovibrational ground state $\ket{X^1\Sigma, v=0, N=0}$ for the average of all sites is shown on the left, while the data on each individual site is shown on the right. The one-way efficiency averaged over all sites is 86(5)\%. We observe a dephasing time of 5.8(7)~\us, which is consistent with scattering loss we expect from the intermediate state. 

One of the advantages of forming molecules in optical tweezer arrays from individual atoms is that the quantum state control we have over the atoms is mapped to the resulting molecules. In particular, the motional state of the resulting molecule is determined by the motional state of the atom pair we start out in. We estimate that $>65(3)\%$ of the rovibrational ground state molecules we form are in the motional ground state of the optical tweezer \cite{Cairncross2021}, limited by heating during the merge process and additional heating processes during the Raman transfer. This constitutes, to the best of our knowledge, the first fully quantum-state-controlled array of rovibrational polar ground state molecules trapped in optical tweezers. 

\section{Conclusion}

In this paper, we presented results on creating an array of fully quantum-state controlled rovibrational ground state polar NaCs molecules in optical tweezers. In the present approach, we used an AOD to create a dual species array of up to 10 Na+Cs atom pairs, and achieved an array of 5 rovibrational ground state NaCs molecules starting with 5 Na+Cs atom pairs. With future implementation of \emph{in situ} rearrangement of atoms and molecules, this presents a starting point for scaling ultracold molecules associated from individual atoms to even larger system sizes \cite{Endres2016, Barredo2016, Sheng2021}. Recent work has demonstrated magic trapping of Na on the D1 transition~\cite{Mujahid2021}, which would enable trapping without the need for switching of the lasers and thus less laser power, and would open possibilities to scale the system size further. In scaling to higher dimensional arrays, spatial light modulators may be used to allow flexible trap geometries and eliminate frequency beating issues encountered in the present work with AOD's Furthermore, spectroscopy work is in progress to identify an intermediate state with a narrower linewidth (5-10~MHz is expected based on the atomic linewidth and other bi-alkali species~\cite{Bause2021},
compared to the \roughly 120MHz observed in the present intermediate state of choice), which would allow higher and more robust molecule transfer efficiency in a larger array. With their large dipole moments, NaCs molecules are poised to generate strong dipole-dipole interactions \cite{Ni2018}. The exquisite control over individual molecules in this platform opens exciting new opportunities in quantum science applications harnessing the rich properties of molecules. 

\textit{Note added -} During the preparation of this manuscript, we became aware of separate but related works detecting over 100 Feshbach molecules at single-molecule resolution in an optical lattice \cite{Rosenberg2021} and a dual species optical tweezer array of Rb and Cs atoms \cite{Singh2021}.

\section{Acknowledgments}
We thank Yu Wang and Gabriel Patenotte for experimental assistance. This work is supported by AFOSR(FA9550-19-1-0089),  AFOSR MURI (FA9550-20-1-0323), and NSF (PHY-2110225). J. T. Z. was supported by a National Defense Science and Engineering Graduate Fellowship. K. W. is supported by a National Science Foundation Graduate Research Fellowship.

\vspace{20pt}

\bibliographystyle{iopart-num}
\bibliography{references}

\end{document}